# MASS DISTRIBUTIONS OF CLUSTERS FROM GRAVITATIONAL MAGNIFICATION

Tom Broadhurst
Physics and Astronomy, The Johns Hopkins University, Baltimore USA
tjb@skysrv.pha.jhu.edu

## ABSTRACT

Lensing in the context of rich clusters is normally quantified from small image distortions, yielding a relative mass distribution in the limit of weak lensing. Here we show the magnification effect of lensing can also be mapped over a cluster, resulting in absolute mass determinations for the weak limit. Furthermore, given both magnification and distortion measurements, the mass distribution may be constrained in the strong regime. Methods for obtaining the magnification using spectroscopic and/or photometric information are discussed, for object detection within a fixed isophote or to a given flux limit. A map of the magnification around A1689 is constructed from the observed depletion of background red galaxy counts.

## 1. Introduction

A large scale measurement of cluster mass profiles provides information on the formation history of structure - a problem inexorably bound up with the thermal properties of the dark matter and the cosmological model. Traditionally cluster masses have been estimated from line of sight velocity dispersions and X-ray properties, indicating a density parameter of only $\Omega = 0.2$ if the M/L of cluster galaxies are representative. These methods, however, are limited in two respects. Firstly, they assume spherical symmetry and that the gas or galaxies are in hydrostatic equilibrium, yet most if not all clusters show substructure (White, Briel & Henry 1994) in accord with idea that clusters undergo continuous hierarchical merging. Secondly, at best these methods can only hope to constrain the mass interior the the radius inside which the galaxies or gas lie; they are blind to mass in an extended halo. Yet it is just such haloes that provide the most natural way to reconcile small scale measurements of $\Omega$ with the expectation of an overall $\Omega = 1$ on theoretical grounds.

A powerful new technique is to measure the gravitational distortion of background galaxies by the gravitational field of galaxy clusters. Rich clusters are found to form very strong lenses and now provide new detail on the shape of the dark matter profiles (Tyson et al 1990, Bonnet et al 1994, Fahlman et al 1994, Smail et al 1994). In principle cluster lensing also allows mass profiles to be measured out to several Mpc, if sufficiently accurate photometry is available, providing by far the cleanest way to measure the mass of clusters.

Distortions alone provide a relative mass distribution, being sensitive to gradients of the mass distribution. However, we show here how an absolute mass measurement can be obtained via the magnification of the background. We describe methods for detecting the magnification and apply one simple method to new photometry of A1689.

## 2. Gravitational Magnification

The importance of measuring the magnification follows from considering the effect of lensing on a small area of sky beyond the lens. An extended image is in general stretched asymetrically, increasing its size and changing its shape. Any anisotropy of lens mass distribution about a given image produces a change in its shape and a corresponding magnification. This effect or shear, $\gamma$, is non-local and can be considered separately from the local convergence, $\kappa$, which is a purely magnifying effect due to the focusing by matter local to the image. The convergence relates directly to the mass distribution and is therefore the quantity of interest, but is not directly observable, being responsible for only part of the overall magnification. However, because the observable quantities of magnification and ellipticity, depend differently on the convergence and shear it is possible to isolate $\kappa$, as follows.

An image is stretched by factors $(1-\kappa-\gamma)^{-1}$ and $(1-\kappa+\gamma)^{-1}$, along orthogonal axes defined by the local shear direction, generating an observed ellipticity, e (=b/a), of an initially circular source at angular coordinate $\theta$:

$$e(\theta) = \frac{1-\kappa-\gamma}{1-\kappa+\gamma}. \tag{1}$$

The total magnification, $A(\theta)$, which is just the ratio of the lensed to unlensed area becomes,

$$A(\theta) = ((1-\kappa)^2 - \gamma^2)^{-1}. \tag{2}$$

(Young et al 1981), so that the convergence is related to these observables via,

$$\kappa(\theta) = 1 - \frac{1+e}{2\sqrt{Ae}}. \tag{3}$$

Strictly, $A(\theta)$, and therefore $\kappa(\theta)$ depend on redshift, asymptoting to a maximum for sources at infinity, so that the convergence relates geometrically to the surface density, $\Sigma(\theta)$, with knowledge of the lens and source distances so that,

$$\Sigma(\theta) = \kappa(\theta) \frac{c^2}{4\pi} \frac{D_s}{D_d D_{ls}}. \tag{4}$$

which simplifies when expressed in terms of the asymptotic value of the convergence $\kappa_\infty$, as then the redshift dependence can be separated, becoming,

$$\Sigma(\theta) = \kappa_\infty 10^{14.44} h M_\odot Mpc^{-2} \frac{(1+z_L)^2}{\sqrt{(1+z_L)}-1}. \tag{5}$$

(Broadhurst, Taylor and Peacock 1994). An issue we shall return to later is the dependence of the derived $\Sigma(\theta)$ on the n(z) of the background sources which because of the magnification is deeper for larger A, to a given flux limit, and if unaccounted for leads to an overestimate of $\Sigma(\theta)$.

The above is only useful for circular sources and in practice must be generalised to cover the more realistic case of images with random initial ellipticities which bias the estimate of $\kappa$. However, this exercise demonstrates the basic behaviour in terms of observables and allows us to explore limiting cases of interest.

## 2.1 Weak Limit

In the general weak limit the convergence can be determined from the magnification alone (eqn 3):

$$\kappa(z) = (A(z) - 1)/2 \qquad (\kappa << 1, \gamma << 1). \tag{6}$$

This is interesting as it shows how, in principle, information on the mass distribution can be obtained independently of the image distortions, hence improved S/N results from combining this information with the detections obtained by the 'classic' Kaiser & Squires (1992) technique for weak distortions.

Another exciting possibility is the extension of mass determinations into the strong regime, along the lines of eqn 3, since $\kappa$ is a function of e and A. So far this has not been possible using only image distortions, as the derived surface density is subject to an unknown transform (Schneider & Seitz 1994, Kaiser 1994).

## 3. Measuring Magnification

Magnification as an observable effect has recently been discussed in the context of rich clusters, as an alternative way of measuring cluster masses (Broadhurst, Taylor and Peacock 1990, BTP). We describe here methods developed by BTP and other ideas, paying attention to the distinction between isophotal and total measures of galaxy magnitudes. This distinction is important as lensing preserves surface brightness, so that in principle if galaxies can be measured to a fixed isophote, then the magnification is just the change in area (of individual galaxies or in the surface density of images) relative to the unlensed sky. In practice defining such an isophote requires very high spatial resolution and this is not yet possible even with HST, given the small sizes of faint galaxies, as described below. However, with good estimates of total magnitudes the magnification can be obtained reliably by its effect on the background surface density of galaxies.

## 3.1 Areal Sizes

The ideal selection of images for measuring magnification would be purely isophotal because of surface brightness conservation. In principle then, the ratio of images areas measured within a given isophote in lensed sky and unlensed comparison sky is the magnification. The difficulty of making this simple test is obvious from examining deep HST images of random sky areas. These images show clearly the intrinsic distribution of image sizes and surface brightness is very broad, extending to very small sizes so that many objects are only partially resolved even with 0.1″sampling (see Kaiser,Squires and Broadhurst 1994, KSB hereafter). This has interesting implications for galaxy formation but a detrimental effect on the ability to set a common isophote for measuring the magnification and certainly means that ground based measures of small lens induced changes in shape are more strongly circularised by the seeing than suspected, (KSB).

## 3.2 Galaxy Counts

A more practical approach is to measure the effect of magnification from the change in surface density of galaxies on the sky. For the ideal case of galaxy selection to a fixed isophotal threshold, the ratio of field and cluster background counts is equal to the magnification. Again in practice we don't yet have such data and must consider the effects of seeing on intrinsically small faint galaxies which make up a good fraction of the background. In the case of ground based photometry it is best to attempt a total magnitude to some reliable limiting flux, which is approximately achieved using a large enough aperture on high S/N images. In this case the counts are depressed equal to the magnification but countered by objects brightened up above the flux limit, so that the relation between lensed an unlensed counts is:

$$N(m)' = N(m)_o A(z)^{2.5s-1}. \qquad (7)$$

where $s = dlogN(m)/dm$ is the intrinsic slope of the number counts. The blue selected galaxy counts are no use for this test as s=0.4, as lensing leaves the counts unaltered (eqn 7). However the reddest galaxies do have a very flat slope $s \approx 0.15$ (see section 4), due to a combination of k-correction and evolution, so that a significant count *depletion* is expected from the areal expansion, $A^{-1}$, which dominates. In practice the galaxy selection will fall between purely isophotal and total magnitudes requiring the curve of growth distribution of the images to be accounted for in an accurate estimate of the magnification but at bright enough magnitudes this is a negligible uncertainty (KSB).

Of course, the counts are subject to fluctuations from the small scale projected clustering of background galaxies and pure shot noise, translating into a limiting 'resolution' for detection of a desired amplification in a given patch of sky behind the lens. For example, at B=26 the clustering fluctuations in the counts are equal to the poisson noise in 1'areas, where the mean count is $10^2$ producing 20% fluctuations in surface density. Going fainter with HST or large ground based telescopes can reduce the shot noise considerably but a residual $\approx 5\%$ clustering or minimum magnification of 15% is the best that can be hoped for on this scale. For radial measurement of the mass, clustering can be smoothed over on large angles since the counts increase at r$^2$, so that the S/N of the magnification measurement is roughly independent of radius (for an isothermal profile). This technique is applied below to photometry of A1689.

### 3.3 Fluctuation Independent Galaxy Counts via Redshifts

Considering the behaviour of the number counts as a function of redshift N(m,z) leads to a clustering independent measure of the amplification but at the cost of requiring redshift information for samples of a few hundred redshifts (BTP). The idea is to make use of the break away from the powerlaw of the luminosity function at the bright end, so that the shape of the distribution of N(m,z) is altered compared to a similarly selected control field sample, and *independent* of density fluctuations (BTP). The probability distribution is constructed:

$$P(m,z) = \frac{N(m + 2.5 log A(z), z)}{\int N(m + 2.5 log A(z), z) dm}. \qquad (8)$$

which is normalised in each redshift bin to the mean field value to remove the uncertain density dependence. BTP have shown that the mass of a cluster can be constrained out to r $\approx$ 1Mpc for the massive clusters (average $\kappa_\infty = 0.05$ over this radius) with only $\approx 300$ redshifts. The gross redshift distribution is altered by the magnification since at every redshift one sees further down the luminosity function for a flux limited sample so that the relationship between the lensed and unlensed N(z) becomes,

$$N(z)' = N(z)_o A^{\beta(z)-1}. \tag{9}$$

Where $\beta = -dlog\Phi(L_{lim}(z), z)/dlogL(z)$, the effective index of the luminosity function at the lower limiting luminosity $L_{lim}$ corresponding to the flux limit. This index increases with redshift in a flux limited sample because at high redshift where the flux limit falls on the steep part of the luminosity function the observed number is increased relative to an unlensed sample, but at lower redshift, just behind the lens, N(z) is suppressed by $\approx A^{-1}$, since the flux limit falls on the flat part of the luminosity function. This effect leads to a stretching of the N(z) to higher redshift and hence a downward correction to the measure of $\kappa$ based only on projected data (BTP).

In general this method can be thought of as an extension to the current distortions approach by the addition of redshift information for the faint images. In the near future we may expect redshifts for many thousands of faint background galaxies (e.g. the HET project now underway, Hill et al 1994) and hence we may realistically hope to establish the form of $A(z)$ over the surface of a cluster.

### 3.4 Standard Candles

Given internal velocity measurements, every galaxy is a rough standard candle via the L-$\Delta V$ relations for discs and early type galaxies. Since the velocities are unaffected by lensing, then at fixed $\Delta V$ any shift in L is just the magnification. Unfortunately, internal velocity measurements are not yet practical at cosmological distances and in any case the intrinsic dispersion in the relation limits the 'resolution' of the magnification measure. 10% precision in $A$ requires $3\Delta L/\sqrt{n} \approx 0.1$ or a minimum of 100 accurate internal velocity measurements, so that in order to achieve 1'resolution in the magnification measurement, given the surface density of galaxies on the sky translates into 1A spectroscopy down to B=26, which is not going to be easy.

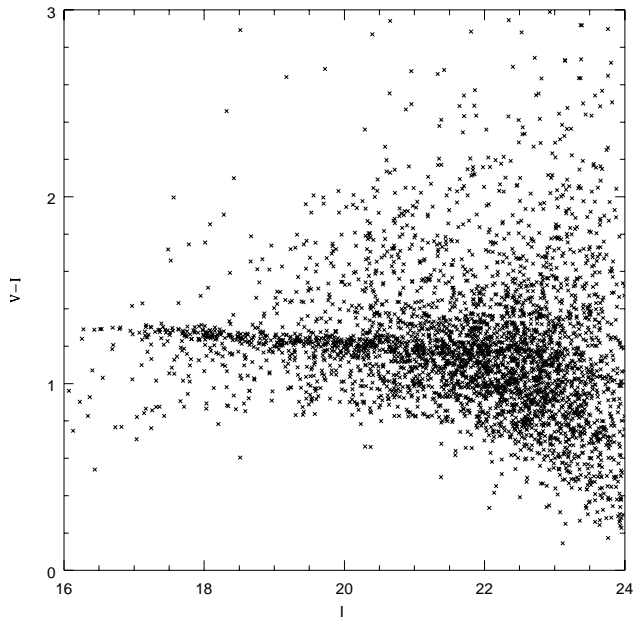

Fig. 1. Colour-Magnitude diagram for A1689 showing the cluster E/S0

sequence as a near horizontal line at V-I=1.3 in this plot. Many redder galaxies are evident and these are taken to be background to the cluster since higher redshifts are required to generate colours redder than the cluster E/S0's.

## 4 Number Counts Behind A1689

Good photometry for A1689 (z=0.18) now exists with which to examine both the magnification and the distortion field. The data covers 70 square arcmins on the cluster and an offset control field in seeing 0.7″, from the ESO NTT 3.6m telescope (Broadhurst et al 1994). The passbands and exposures were chosen such that the cluster E/S0 galaxies would be bluer than a good fraction of the background, requiring much deeper imaging in the bluer passband for detection. For red objects we can be confident they are in the background since redshifts greater than the cluster are required to generate colours redder than an early type at the cluster redshift. The red counts have the further important advantage that they are relatively flat so that the magnification is dominated by the areal expansion (eqn 7).

The photometry was carried out by Nick Kaiser using a carefully optimised scheme (see KSB for details of procedure and calibration), obtaining near total magnitudes within 3 times the Petrosian radius and accurate image shapes. Figure 1 is the colour magnitude distribution, showing a well defined E/S0 sequence for this cluster, and also the spread of redder background galaxies. The distortion map obtained from the KS technique is discussed by Kaiser et al (1994) showing a $9\sigma$ detection of the cluster and a smooth radial mass profile out to the limit of the data, making it the best detection of a cluster to date by the KS technique.

Here we explore the magnification via the number counts of the background galaxies. The far field red galaxy number counts are observed to have an intrinsically flat slope for $I > 20$, as predicted, $s = 0.15$, and is well fitted by extrapolating the local early type luminosity function to higher redshift (solid curve). Figure 2 shows the far field red counts and the no-evolution model (upper panel) scaled to the area of the field for V-I>1.5, which is 0.2 mag redder than the upper limit of the E/S0 sequence. In contrast the blue counts, V-I<1, have the steep $s = 0.4$ slope which can be reproduced with density evolution at a rate $(1 + z)^3$, as shown in Figure 2 (lower panel). This relative evolution in V-I is well known from evolutionary studies and is most clearly demonstrated by the luminosity functions of Lilly et al (1994) in these passbands.

In measuring the surface density of background objects the coverage of the background by the cluster galaxies is determined. This is easy to quantify by tracing the radial distribution of areal coverage of all the galaxy images and finding the level at which it flattens off to the background level. The difference is then the coverage which is found to be important only in the central 30″ of this cluster due to a large central concentration of cD's. In any case this central region is within the Einstein radius as defined by the prominent giant arc surrounding the center and therefore cleared of background images apart from counter images (see figure 3). The number counts in the central region, r<3′, are shown in Figure 2, for the red and blue galaxies and compared with isothermal model lensed predictions by modifying the count model (eqn 9) for total magnitudes (dotted histograms and curves respectively). Examination of the effects of seeing on this assumption are being investigated by Kaiser & Broadhurst from 2 colour HST data of the 'Medium Deep Survey'). For the red counts a strong effect is observed, well matched by an isothermal mass distribution of 1600km/s and showing the predicted depletion from the areal expansion of the sky by the lens.

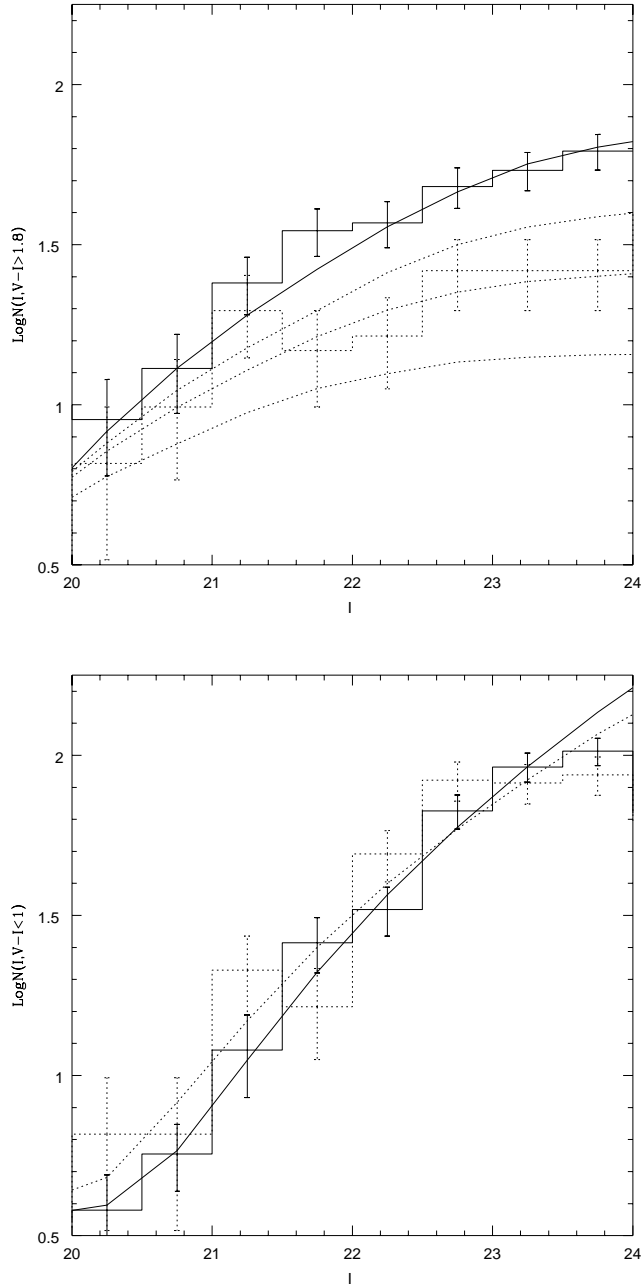

Fig. 2. Comparison of the far field number counts of red galaxies (upper panel solid histogram) and blue galaxies (lower panel solid histogram) with the number-count model predictions for the far field and evolution as found by Lilly et al (solid curves) in V and I. The red and blue counts within a radius of 3′are also shown (corrected for coverage as described in the text) as dotted histograms and compared with the lensed counts with the lensing modified count predictions for isothermal profiles with finite core radius as described in the text. It is clear that the red counts are very sensitive to lensing as expected given their intrinsically flat slope, matching best $v_{11}$ = 1600km/s. This contrasts with the insensitivity of the blue counts which have the invariant dlogN/dm=0.4 slope (lower panel).

For the blue counts no useful constraint is possible as expected, given the invariant s=0.4 count slope, and demonstrated by the small difference between the lensed and unlensed expectation shown in Figure 2 (lower panel), in this case the 1600km/s isothermal model is plotted - dotted curve. The radial depndence of the red count depletion is shown in Figure 3 versus the three isothermal models of figure 2a with a core radius of $0.3 \times \theta_{crit}$ (normalised to the far field). The observed profile is shallower than isothermal, with some evidence of a break in the profile at 4′, although further checks are required. The model profiles show the critical radius as a minimum in the surface density. The increase in density interior to this radius comes from the multiple images and is very sensitive to the core radius. Unfortunately the bright galaxies in the core effectly blot out this interesting region of the sky, in the optical at least. Figure 3 also shows agreement between the minima of the data and the $v_{11}$=1600km/s model at ≈ 50″. This is interesting as it predicts well the observed critical radius. Several giant arcs are seen around the center in our data, ranging in radius from 45-65″(Broadhurst et al 1995). Note, the dispersion in radius of these giant arcs is a combination of split images straddling the critical curve and also the range in redshift of the background sources. This agreement is an encouraging check on the consistency of the method.

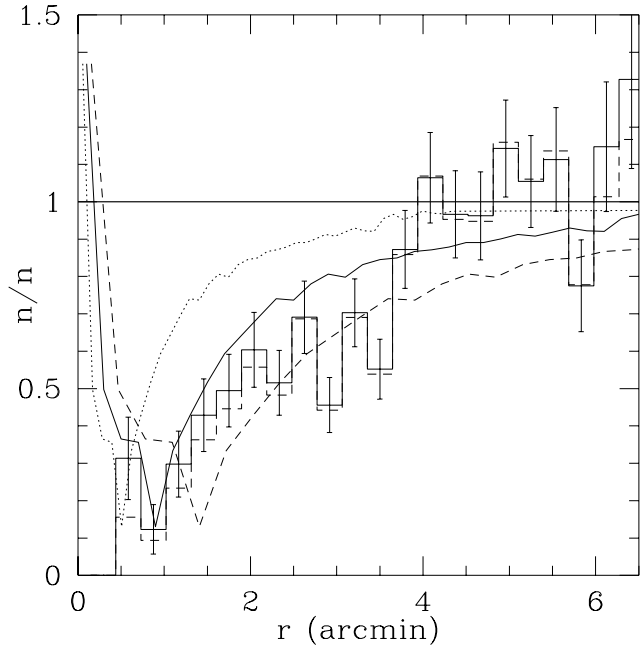

Fig. 3. Radial counts (unsmoothed) normalised by area and corrected upward for coverage by the cluster galaxies. Dashed histogram shows the counts without the correction for by the coverage mask, demonstrating the small relative coverage by the cluster. The curves are the expected form of the radial counts for the isothermal cases, with $v_{11} = 1200, 1600, 2000$km/s (dotted, solid and dashed curves respectively) showing the turn up interior to the critical radius from split images (core radius=$0.3\theta_{crit}$). The observed profile is shallower than isothermal and may show a break at 4′, requiring confirmation.

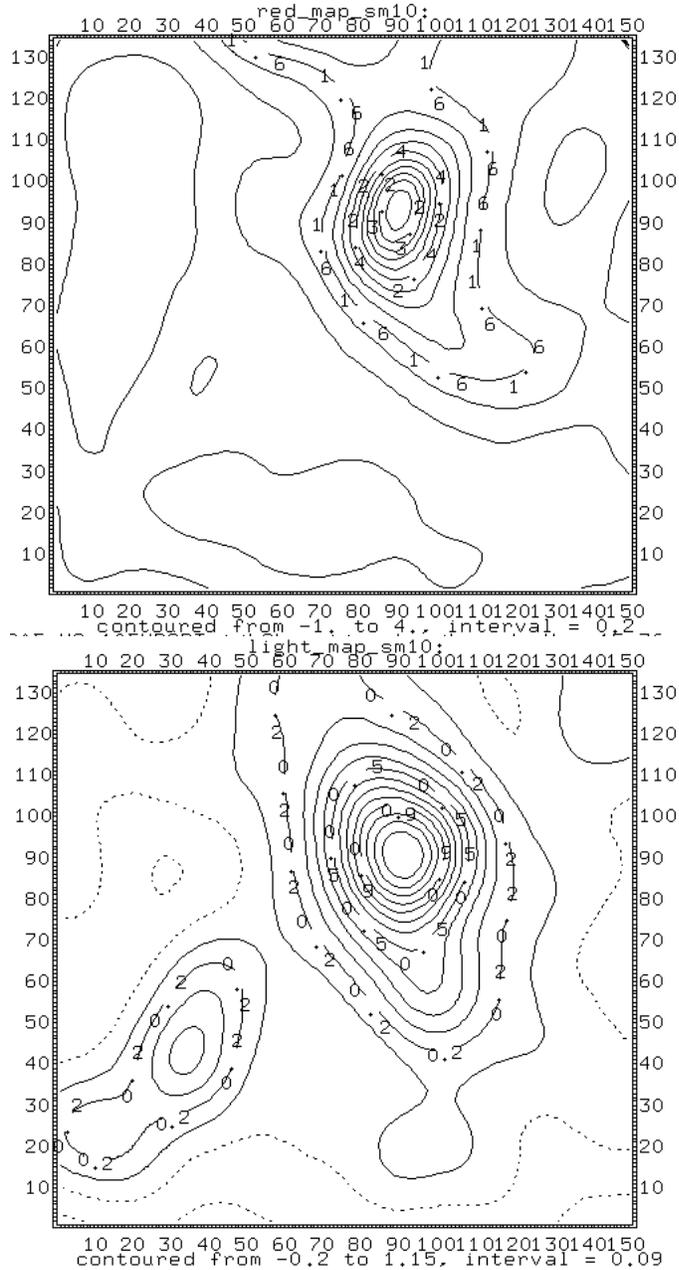

Fig. 4. Comparison of the amplification map over the full 8′ of the data (upper panel) for comparison with the cluster light map (lower panel) at the same resolution, (30″ gaussian smoothing). The agreement in position of the peaks of light and amplification is very good (Small light peak in lower left is due to the halos of 2 bright stars not fully dealt with as yet). Conversion of this to a mass map requires incorporating the image distortions via eqn 3 in the strong limit.

A simple direct estimate of the mass over the area covered by the data (0.8Mpc/h on a side) using the the simple weak limit relation derived above $\kappa = (A(z)-1)/2$ applied to $A(z)$ measured via eqn 9, yields a mean surface density of $3 \approx 10^{15} h M_\odot Mpc^{-2}$. However, more work is required to obtain sensible constraints in the inner strongly lensed regions as the shear is strong here generating magnification in addition to that of the local convergence (eqn 3). requiring use of the ellipticity information contained in the image distortion map of this cluster.

Interesting detail is seen the map of the amplification distribution compared to that of the light distributions as shown in Figure 4. Both maps are at the same 30″resolution (scale of the maps is 8′on a side). Data on this cluster is also good enough to produce a detailed distortion map (Kaiser et al 1994), and hence we will be able to reconstruct fully the mass distribution over the surface of this cluster by combining measurements of ellipticity and magnification.

## CONCLUSIONS

In conclusion, it would appear that the magnification effect by the cluster A1689 is clearly detected, depressing the surface density of the background red galaxies. The detection is strong enough to allow a crude map of the magnification. Further work is required to reconstruct the mass distribution by combining this information with the image distortion field of the cluster. At that point we can expect to define an accurate mass distribution and also a map of the mass-to-light ratio, in both the strong and weak limits.

## ACKNOWLEDGEMENTS

Many thanks to Steph Côté, Alex Szalay, Ed Turner and especially Nick Kaiser for helpful discussions.